\documentstyle{mn}
\input epsf

\begin{document}

\newcommand{\dD}{\delta_{\rm D}}
\newcommand{\rhob}{\bar{\rho}}
\newcommand{\nb}{\bar n}
\newcommand{\bx}{{\bf x}}
\newcommand{\br}{{\bf r}}
\newcommand{\brp}{{\bf r'}}
\newcommand{\bR}{{\bf R}}
\newcommand{\bRp}{{\bf R'}}
\newcommand{\bk}{{\bf k}}
\newcommand{\bkp}{{\bf k'}}
\newcommand{\hMpc}{$h^{-1}\, {\rm Mpc}\, $}
\newcommand{\Mpch}{$h\, {\rm Mpc}^{-1}\, $}

\journal{Preprint-98}

\title{The Growth of Correlations in the Matter Power Spectrum}

\author[A.Meiksin and M. White]{A. Meiksin${}^{1}$ and Martin White${}^{2}$\\
${}^{1}$Institute for Astronomy, University of Edinburgh,\\
Blackford Hill, Edinburgh EH9 3HJ, UK\\
${}^{2}$Departments of Physics and Astronomy, University of Illinois\\
at Urbana-Champaign, Urbana, IL 61801-3080, USA}
\pubyear{1998}

\maketitle

\begin{abstract}
We find statistically significant correlations in the cosmological
matter power spectrum over the full range of observable scales.
While the correlations between individual modes are weak, the band-averaged
power spectrum shows strong non-trivial correlations. The correlations are
significant when the modes in either one or both bands are in the non-linear
regime, and approach 100\% for pairs of bands in which all the modes are
non-linear. The correlations are weaker, but not absent, when computed in
redshift space. Since estimates of the power spectrum from galaxy surveys
require band-averaging, the correlations must be taken into account when
comparing a measured power spectrum with theoretical models.

\end{abstract}


\section{Introduction}

In our current paradigm for cosmological structure formation, initially small,
primordial density perturbations grow through gravitational instability to
form the large-scale structures that we see in the distribution of galaxies
today.  The most popular assumption is that the primordial fluctuations were
distributed according to a homogeneous Gaussian random process.  Under this
assumption all of the statistical information is encoded in the power
spectrum.
The assumption of initially Gaussian modes is a prediction of the simplest
models of inflation, and has received some limited observational support
through measurements of the CMB \cite{KBBGHSW,Heavens} and large-scale
structure \cite{BSDFYH,Gaz,NusDekYah,FKP,Colley}.
However, it is known that as the perturbations grow and become non-linear,
the modes become coupled.
In this paper, we study this process quantitatively using $N$-body simulations.

Since it is the lowest order non-vanishing statistical description of the
density or galaxy field, the power spectrum has been at the focus of much
recent attention.  Many galaxy surveys have attempted to determine the nature
of the power spectrum in two and three dimensions, and two massive surveys,
the Anglo-Australian Two Degree Field
(2dF\footnote{http://meteor.anu.edu.au/$\sim$colless/2dF})
and the Sloan Digital Sky Survey
(SDSS\footnote{http://www.astro.princeton.edu/BBOOK})
are moving the subject toward high-precision measurements.
Several methods for performing detailed comparisons between a measured power
spectrum from a galaxy survey and a theoretical power spectrum have been
devised (see Tegmark et al. 1998 for a review).
Critical to assessing the accuracy of the measurements is an evaluation
of the statistical properties of the power spectrum.
To date most estimates of the precision to which the power spectrum may be
measured have assumed that the Fourier components of the density fluctuations
are still distributed as a Gaussian random process.
In that case the standard deviation of the power spectrum is determined by
the value of the power spectrum.
We expect this assumption to be valid at early times and on sufficiently large
scales for models with Gaussian initial conditions.
The question of how large is ``sufficiently large'' requires a detailed
calculation.

We show here that non-linear clustering gives rise to significant correlations
in the band-averaged power spectrum at different scales in the regime
of observational interest. The correlations increase with increasing
(spatial) frequency, reaching levels near 100\% when the perturbations
are non-linear. Moreover, the correlations require the modes in only a single
band to be non-linear:\ we find significant correlations between high and
low frequencies even if the modes in the low frequency band are still in the
linear regime. A consequence of the mode coupling is an increase in the
dispersion of the power spectrum, suggesting that estimates of the
accuracy to which the power spectrum may be measured should be revisited.

The outline of the paper is as follows: in \S\ref{sec:theory} we discuss
the expectations for correlations in the power spectrum between modes.
In \S\ref{sec:nbody} we describe our numerical experiments, while in
\S\ref{sec:stats} we discuss the statistical properties of the correlations.
We finish in \S\ref{sec:conclusions} with some comments on the impact of
our results for future surveys designed to measure the matter power spectrum.

\section{Correlations in the Power Spectrum} \label{sec:theory}

The density contrast is defined by $\delta(\br)=[\rho(\br)-\rhob]/\rhob$,
where $\rho(\br)$ is the matter density at position $\br$, and $\rhob$ is
the mean density.
We consider a large spatial volume $V$ and impose periodic boundary conditions,
thus $\br$ is continuous but its Fourier conjugate $\bk$ is quantized.
One can take the continuum limit by replacing $V^{-1}\sum_\bk$ with
$\int d^3k/(2\pi)^3$.
Our Fourier transform convention has
\begin{equation}
  \delta_\bk=V^{-1}\int\, d^3r\ \delta(\br) \exp[i\bk\cdot\br]
  \qquad ,
\end{equation}
and
\begin{equation}
  \delta(\br)={V\over (2\pi)^3}\int\, d^3k\ \delta_\bk \exp[-i\bk\cdot\br]
\end{equation}
both dimensionless.
We define the power spectrum as
$P(k)=V\left\langle |\delta_k|^2\right\rangle-1/\nb$, where the angled
brackets $\langle\cdots\rangle$ denote an ensemble (and Poisson) average,
and the shot noise term $1/\nb$ has been subtracted. Here, $\nb$ is the
mean number density of objects used to measure the power spectrum.
The power spectrum is related to the 2-pt spatial correlation function
$\xi(r)$ by
\begin{equation}
  P(k)=\int\, d^3r\ \xi(r)\exp[i\bk\cdot\br]
  \qquad .
\end{equation}
We denote our (unbiased) estimate of the power spectrum at $\bk$, which we
obtain from the simulations, as
\begin{equation}
  \widehat{P}(\bk)\equiv V\left| \delta_\bk \right|^2-1/\nb
  \qquad .
\end{equation}
Typically we reduce the scatter in this quantity by averaging over
a thin shell in $\bk$-space with $|\bk|\simeq k$.

The covariance between two random variables $x$ and $y$ is defined as
${\rm cov}[x,y]=\langle xy\rangle-\langle x\rangle\langle y\rangle$.
The variance is ${\rm var}(x)={\rm cov}[x,x]$.
The correlation between $x$ and $y$ is defined as
$\rho(x,y)={\rm cov}[x,y]/ [{\rm var}(x){\rm var}(y)]^{1/2}$.
The covariance of our power spectrum estimator, $\widehat{P}(\bk)$,
depends on the 2-pt function (power spectrum), the 3-point function
(bi-spectrum) and the 4-point function (tri-spectrum).
We obtain, to leading order in $(\nb V)^{-1}$ as $\nb V\to\infty$
(Peebles 1980, \S36),
\begin{eqnarray}
{\rm cov}[\widehat{P}(\bk),\widehat{P}(\bkp)] &=&
  \left[ P(k)+\nb^{-1} \right]^2
  \left(\delta_{\bk,\bkp}+\delta_{\bk,-\bkp}\right) \nonumber \\
&+& T(\bk,-\bk,\bkp,-\bkp)/V
\label{eqn:leading}
\end{eqnarray}
with $\delta_{\bk,\bkp}=1$ for $\bk=\bkp$, and $0$ otherwise, and
where we have written the tri-spectrum as $T$. The tri-spectrum is defined as
the Fourier transform of the 4-point function $\eta$ according to
\begin{equation}
 T(\bk_1,\bk_2,\bk_3,\bk_4)=
\label{eqn:trispectrum}
\end{equation}
\begin{eqnarray}
\frac{1}{V}\int d^3x_1\, d^3x_2\, d^3x_3\, d^3x_4\ \eta({\bf x}_1,{\bf x}_2,
{\bf x}_3,{\bf x}_4) \times \nonumber \\
\exp[i\bk_1\cdot{\bf x_1}+i\bk_2\cdot{\bf x_2}+i\bk_3\cdot{\bf x_3}
+i\bk_4\cdot{\bf x_4}].  \nonumber
\end{eqnarray}
The first term in Eq.~(\ref{eqn:leading}) is the usual result for Gaussian
fluctuations, viz ${\rm var}[x^2]=2\langle x^2\rangle^2$ for a real variable
$x$. Because $\delta_\bk$ is complex with uncorrelated real and imaginary
parts,
$P=2\langle [{\rm Re}(\delta_\bk)]^2\rangle=
 2\langle[{\rm Im}(\delta_\bk)]^2\rangle$,
and the Gaussian contribution to the error on $P$ becomes equivalent to the
total power spectrum: the sum of the signal and the noise.
Only the tri-spectrum contributes to the off-diagonal ($k\neq k'$) elements
of the covariance.

The terms sub-dominant in $\nb V$, all shot noise terms, are also
straightforward to derive.  There is a constant term $(V\nb^3)^{-1}$,
contributions from the 2-pt function
\begin{equation}
{1\over \nb^2 V}\left[
  P(|\bk+\bkp|) + P(|\bk-\bkp|) + 2P(k) + 2P(k')\right]
\end{equation}
and the 3-pt function
\begin{eqnarray}
{1\over \nb V} &&
\left[ B(\bk,-\bk,{\bf 0})+B({\bf 0},\bkp,-\bkp)+\right. \nonumber \\
&& B(\bk+\bkp,-\bk,-\bkp)+B(\bk-\bkp,-\bk,\bkp)+ \nonumber \\
&& \left. B(\bk,\bkp-\bk,-\bkp)+B(\bk,-\bk-\bkp,\bkp)\right]
\end{eqnarray}
where the bi-spectrum is defined as the Fourier transform of the 3-pt function
$\zeta$
\begin{eqnarray}
B(\bk_1,\bk_2,\bk_3)&=&\frac{1}{V}\int d^3x_1\, d^3x_2\, d^3x_3\
\zeta({\bf x}_1,{\bf x}_2,{\bf x}_3) \times \nonumber \\
&&  \exp\left[ i\bk_1\cdot{\bf x_1}+i\bk_2\cdot{\bf x_2}
+i\bk_3\cdot{\bf x_3} \right] \quad .
\end{eqnarray}

Finally we consider the result for band-averaged estimates of $P(k)$.
Denote by $\widehat{P}_i$ our estimator $\widehat{P}(\bk)$ averaged over
a set of $N_i$ $\bk_{i,\alpha}$ with $|\bk_{i,\alpha}|\simeq k_i$ for
$\alpha=1$ to $N_i$.  Straightforwardly
\begin{equation}
  {\rm cov}\left[ \widehat{P}_i , \widehat{P}_j \right] =
  {1\over N_i N_j} \sum_{\alpha=1}^{N_i} \sum_{\beta=1}^{N_j}\ {\rm cov}
  \left[ \widehat{P}(\bk_{i,\alpha}), \widehat{P}(\bk_{j,\beta}) \right]
  \qquad .
\label{eqn:baPc}
\end{equation}
Even when the bands $i$ and $j$ are disjoint, we shall see that the
band-averaging introduces non-trivial {\it correlations} in the power spectrum.
These correlations are given by the tri-spectrum averaged over the
configurations in the shell.
In principle one could apply different weights to the modes in the band,
but we have not pursued this line of inquiry.  It is traditional to merely
average over the directions in a $k$-shell and this is what we have chosen
to do in the simulations.

\section{The Growth of Correlations} \label{sec:nbody}

Gravitational perturbation theory (PT) has given some insight into the growth of
clustering beyond the linear regime, and in particular the effect of
clustering on the evolution of the power spectrum, the bi-spectrum, and the
tri-spectrum \cite{P80,Jus81,Vish83,Fry84}. Perturbation theory shows that
the variance in the power spectrum will grow for modes that are still
well in the linear regime. In addition, PT shows that the non-linear growth of
density fluctuations will give rise to a tri-spectrum \cite{Fry84}.
The form the tri-spectrum takes in PT involves multiples of the first order
power spectrum and no powers of $V$. While this is shown explicitly by Fry
to the lowest non-vanishing order in the tri-spectrum, it is straightfoward
to extend the result to all orders. Consequently, the contribution of the
tri-spectrum to the power spectrum covariance between any off-diagonal pair
of $k$-modes is of order $1/V$ in PT, and so vanishes in the large volume limit
according to equation~(\ref{eqn:leading}).
Fan \& Bardeen~\shortcite{FB95} have argued that the power covariance matrix
should remain diagonal even in the non-linear regime.

Nonetheless, as we describe below, {\it correlations} in the
{\it band-averaged\/} power spectrum may
remain even in the limit $V\rightarrow\infty$.
This is a consequence of the decrease in the variance of the power spectrum
with the increased number of modes in the band.
Since the number of modes in a band is proportional to $V$, the $1/V$
suppression of the covariance between modes cancels in the expression for the
correlation $\rho$, leading to a finite value even as $V\to\infty$.
We investigate the development of the correlations in the non-linear regime
using $N$-body simulations.

We remark that for a {\it finite}-volume survey, we would expect the presence
of both covariances and correlations between individual modes according to
equation~(\ref{eqn:leading}) and the associated shot-noise terms.

Using a particle-mesh (PM) code, described in detail in
Meiksin, White \& Peacock~\shortcite{MWP99}, we have performed a series of
several thousand realizations of the evolution of clustering in a
$\Lambda$CDM universe.
The parameters chosen are $\Omega_0=0.4$, $\Omega_\Lambda=0.6$, $h=0.65$,
$\Omega_{\rm B}h^2=0.03$, and $n=1.030$, where $n$ is the primordial spectral
index.
The power spectrum is COBE-normalized using the method of
Bunn \& White~\shortcite{BW97}.
The slight tilt has been chosen to reproduce the abundance of rich clusters
of galaxies today \cite{WEF93}, specifically the {\it rms\/} mass fluctuation
in an 8~\hMpc sphere is $\sigma_8\simeq 0.92$.
The non-linear scale, where the variance becomes unity, is
$k_{\rm nl}=0.19\,h\,{\rm Mpc}^{-1}$.
Tests involving the higher order moments indicate that the code should
accurately reproduce the bi- and tri-spectrum on the scales of interest
\cite{Moments}.

\begin{table*}
\caption{Power spectrum correlation matrix for 200~\hMpc simulation. The last
row shows $\sigma=[\Delta^2(k)]^{1/2}$.}
\begin{tabular}{lrrrrrrrrrrrr}
$k$ (\Mpch)   & 0.031 & 0.044 &  0.058 & 0.074 & 0.093 & 0.110 & 0.138 & 0.169 & 0.206 & 0.254 & 0.313 & 0.385 \\
\\
0.031 &  1.000 & -0.017 & 0.023 & 0.024 & 0.042 & 0.154 & 0.176 & 0.188 & 0.224&  0.264 & 0.265 & 0.270 \\
0.044 & -0.017 & 1.000 & 0.001 & 0.024 & 0.056 & 0.076 & 0.118 & 0.180 & 0.165 & 0.228 & 0.234 & 0.227 \\
0.058 &  0.023 & 0.001 & 1.000 & 0.041 & 0.027 & 0.086 & 0.149 & 0.138 & 0.177 & 0.206 & 0.202 & 0.205 \\
0.074 &  0.024 & 0.024 & 0.041 & 1.000 & 0.079 & 0.094 & 0.202 & 0.229 & 0.322 & 0.343 & 0.374 & 0.391 \\
0.093 &  0.042 & 0.056 & 0.027 & 0.079 & 1.000 & 0.028 & 0.085 & 0.177 & 0.193 & 0.261 & 0.259 & 0.262 \\
0.110 &  0.154 & 0.076 & 0.086 & 0.094 & 0.028 & 1.000 & 0.205 & 0.251 & 0.314 & 0.355 & 0.397 & 0.374 \\
0.138 &  0.176 & 0.118 & 0.149 & 0.202 & 0.085 & 0.205 & 1.000 & 0.281 & 0.396 & 0.488 & 0.506 & 0.508 \\
0.169 &  0.188 & 0.180 & 0.138 & 0.229 & 0.177 & 0.251 & 0.281 & 1.000 & 0.484 & 0.606 & 0.618 & 0.633 \\
0.206 &  0.224 & 0.165 & 0.177 & 0.322 & 0.193 & 0.314 & 0.396 & 0.484 & 1.000 & 0.654 & 0.720 & 0.733 \\
0.254 &  0.264 & 0.228 & 0.206 & 0.343 & 0.261 & 0.355 & 0.488 & 0.606 & 0.654 & 1.000 & 0.816 & 0.835 \\
0.313 &  0.265 & 0.234 & 0.202 & 0.374 & 0.259 & 0.397 & 0.506 & 0.618 & 0.720 & 0.816 & 1.000 & 0.902 \\
0.385 &  0.270 & 0.227 & 0.205 & 0.391 & 0.262 & 0.374 & 0.508 & 0.633 & 0.733 & 0.835 & 0.902 & 1.000 \\
\hline
$\sigma$ & 0.17 & 0.25 & 0.33 & 0.44 & 0.55 & 0.62 & 0.76 & 0.91 & 1.07 & 1.29 &
 1.55 & 1.87
\end{tabular}
\label{tab:corr}
\end{table*}

\begin{figure}
\begin{center}
\leavevmode \epsfxsize=3.3in \epsfbox{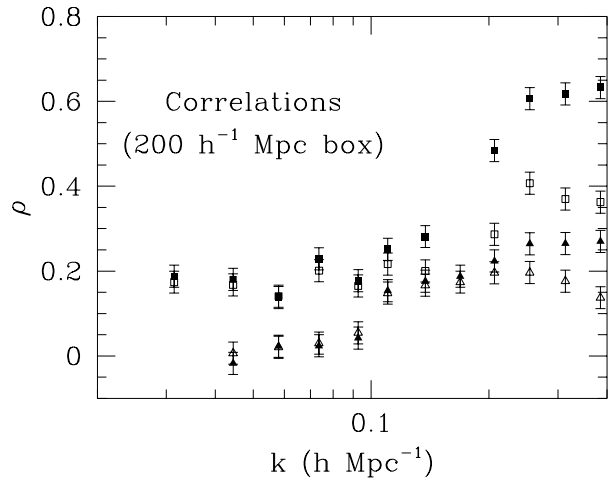}
\end{center}
\caption{The correlations in power between a band centred at $k$ and bands
centred at $k=0.031 h\, {\rm Mpc}^{-1}\,$ (triangles) and
$k=0.17 h\, {\rm Mpc}^{-1}\,$ (squares).
The correlations are shown both in real space (filled) and redshift
space (open). The correlations become significant when the modes in one of
the bands enter the non-linear regime. Redshift space distortions
reduce the correlations on small scales.}
\label{fig:crz}
\end{figure}

All the simulations were run with a $128^3$ force grid and either $128^3$
or $256^3$ particles to isolate the effects of shot noise. Two box sizes were
used, 200~\hMpc and 400~\hMpc, to test for finite volume effects.
For each box size, $N\sim 10^3$ realizations with different Gaussian initial
conditions were evolved from $1+z=20$ to the present. For the $256^3$
simulations, $N\sim 10^2$ realizations were performed. We verified that
starting the simulations at $1+z=30$ did not alter our results. For each
realization the power spectrum was calculated in 20
bins logarithmically spaced in $k$ from the fundamental mode to the
Nyquist frequency of the force grid, as described in Meiksin et al. (1999).
The ensemble averages of the previous section were approximated by an average
over the $N$ realizations.  Convergence of the correlations was typically
obtained with several hundred realizations.

The resulting correlations in the real space power spectrum at $z=0$
for several selected frequency bands from the 200~\hMpc simulations are
shown in Table~\ref{tab:corr}.
For each entry we may assess the likelihood of no correlation using the
Spearman rank order correlation coefficient $r_s$ \cite{NPS}.
The probabilities are $<0.001$ for pairs of bands with both central
frequencies exceeding 0.14~\Mpch.
An estimate of the probability for {\it all\/} the correlations is provided
by the Kendall concordance statistic $W$ \cite{NPS}, which is linearly
related to the average of the $r_s$ values for all the band
pairs.\footnote{The relation between $W$ and the average coefficient $\bar r_s$
for $N_b$ bands is $W=(1-N_b^{-1})\bar r_s +N_b^{-1}$. The correlation
coefficient $r_s$ between two quantities is the product-moment coefficient
computed by replacing the values of the quantities by their ranks.}
For the 12 frequency bands shown in Table~\ref{tab:corr}, we obtain $W=0.33$,
with a vanishingly small probability for obtaining a value so large assuming
no correlations.
Indeed, it is only for correlations for which both bands are confined to
frequencies $k\leq0.093$~\Mpch that the collective probability for
non-correlation is found to exceed 0.001.

We also compute the correlations in redshift space by including the peculiar
velocities of the particles. The correlations are weaker ($W=0.26$ for the 12
bands), but are still highly significant. We expect that the correlations are
weaker in redshift space both because of the reduced power on small scales in
redshift space and because the peculiar velocities introduce extra randomness
which destroys the correlations.

In Figure~\ref{fig:crz}, the correlations are shown between each band
centred at $k$ and the bands centred at $k=0.031$~\Mpch or
$k=0.17~$\Mpch. The error bars indicate one standard deviation
calculated from $\sigma_\rho=N^{-1/2}$, assuming the modes are drawn
from a Gaussian distribution.\footnote{The error is independent of the number
of modes entering the band averages. The errors will be reduced in the
presence of correlations. The error bars shown demonstrate the significance
of the correlations, but do not strictly show the error in the calculated
correlation strength when the correlations are non-vanishing.} In the last row
of Table~\ref{tab:corr}, we show the density fluctuations on each scale $k$, as
defined by $\sigma=[\Delta^2(k)]^{1/2}$, where $\Delta^2(k)=k^3P(k)/(2\pi^2)$.
While the correlations between pairs of low frequency bands, both well in the
linear regime ($k<k_{\rm nl}$), are consistent with zero, the power spectrum
correlation between low frequency bands and high frequency bands
rapidly rises as the frequency of the latter increases. The correlations
involving the higher frequency bands are significantly reduced in
redshift space.

\begin{figure}
\begin{center}
\leavevmode \epsfxsize=3.3in \epsfbox{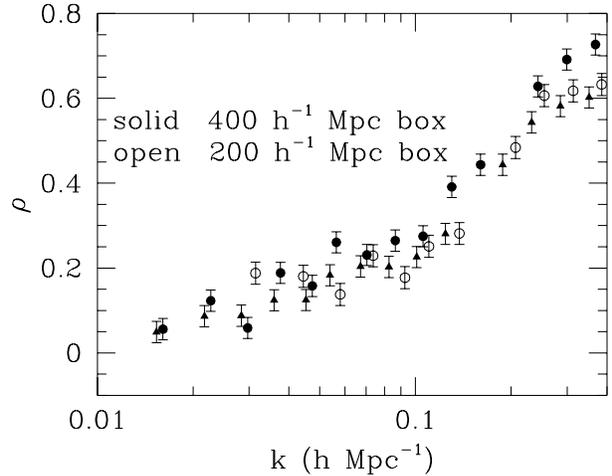}
\end{center}
\caption{The correlations in real space power between the bands centred at $k$
and $k=0.17 h\, {\rm Mpc}^{-1}\,$ (open circles) in the
200 $h^{-1}\, {\rm Mpc}$
simulation. Also shown are the correlations for the
$k=0.16 h\, {\rm Mpc}^{-1}\,$ band (filled triangles) and the
$k=0.19 h\, {\rm Mpc}^{-1}\,$ band (filled circles) in the
400 $h^{-1}\, {\rm Mpc}$ simulation.
The filled points are slightly offset in frequency for clarity. No significant
reduction in the correlations is found in the larger box compared with the
smaller, suggesting that the correlations are not a finite volume effect.}
\label{fig:cb2b4}
\end{figure}

\begin{figure}
\begin{center}
\leavevmode \epsfxsize=3.3in \epsfbox{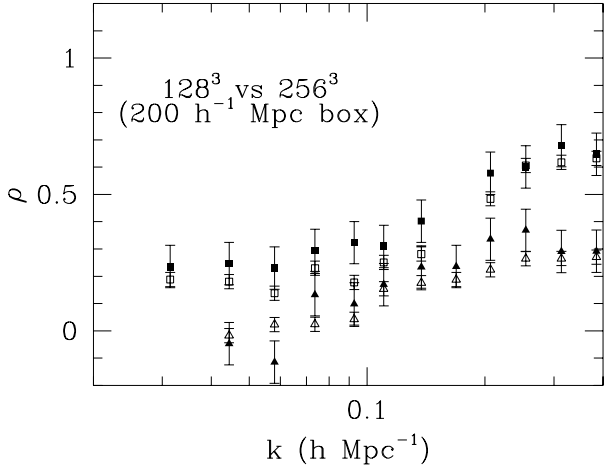}
\end{center}
\caption{The correlations in power between the band centred at $k$ and the
bands centred at $k=0.031 h\, {\rm Mpc}^{-1}\,$ (triangles) and
$k=0.17 h\, {\rm Mpc}^{-1}\,$ (squares).
The correlations are shown for the 200~\hMpc box simulations
for $128^3$ (open) or $256^3$ (filled) particles.}
\label{fig:cshot}
\end{figure}

In Figure~\ref{fig:cb2b4}, we show that no significant difference is found
between the 200~\hMpc and 400~\hMpc box simulations, suggesting that the
correlations have converged and are not due to the finite volume of the
simulations. Similarly, we show in Figure~\ref{fig:cshot} that the
correlations do not decrease as the number of particles increases,
demonstrating that the correlations are not dominated by shot noise
contributions.

\begin{figure}
\begin{center}
\leavevmode \epsfxsize=3.3in \epsfbox{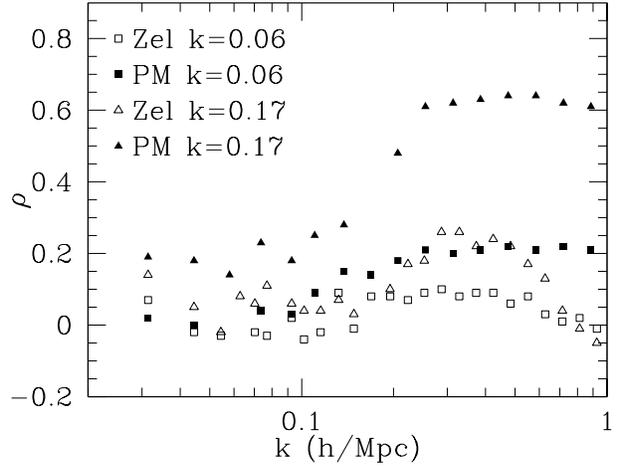}
\end{center}
\caption{The correlations in power between the band centred at $k$ and the
bands centred at $k=0.06 h\, {\rm Mpc}^{-1}\,$ (squares) and
$k=0.17 h\, {\rm Mpc}^{-1}\,$ (triangles) computed using either
the Zel'dovich approximation (open) or the PM code (filled).
While the scatter at low-$k$ is large due to the small number of
modes present, the PM correlations are stronger than the Zel'dovich
correlations.}
\label{fig:cmpzel}
\end{figure}

\begin{figure}
\begin{center}
\leavevmode \epsfxsize=3.3in \epsfbox{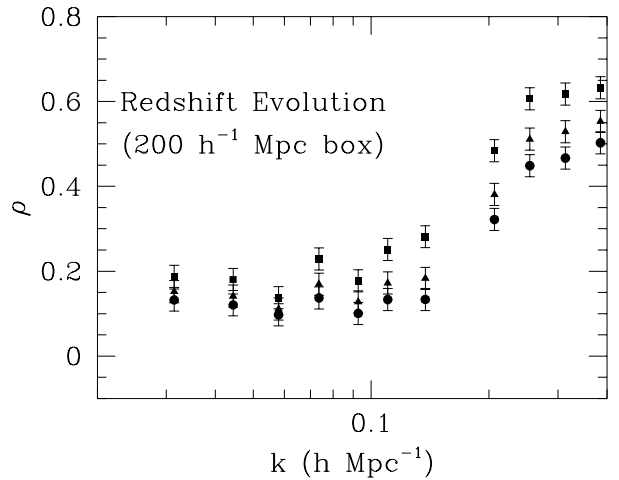}
\end{center}
\caption{The correlations in real space power between the bands centred at $k$
and $k=0.17 h\, {\rm Mpc}^{-1}\,$.
The correlations are shown for 3 different redshifts: $z=0.5$ (circles),
$z=0.3$ (triangles), and $z=0$ (squares). The correlations grow with time.}
\label{fig:czev}
\end{figure}

As an additional test, we performed a similar set of $\sim10^3$ runs
using the Zel'dovich
approximation with the same box sizes and particle numbers as for the PM code.
With the Zel'dovich calculations we could displace the particles from random
positions within the simulation volume, from a regular grid or from random
positions near grid zones. This allowed us to test that our initial
conditions did not contribute spuriously to the correlations. We found,
see Figure~\ref{fig:cmpzel}, that the correlations appeared much more
strongly in the PM runs than the Zel'dovich runs, suggesting that they are
induced by gravitational instability, as shown in Figure~\ref{fig:czev}, and
are not an artifact of our numerical technique. We note, however, that evolving
the particles in the Zel'dovich approximation, even when the particles were
initially placed completely at random within the simulation box, again
resulted in statistically significant correlations.

We have also tested the random number generation
by using two separate generators for 512 Zel'dovich runs each, one based on a
multiplicative congruential generator \cite{NumRec}, and the other based on a
lagged Fibonacci generator \cite{B95}, the periods of both of which
should be ample for our purposes. The results were statistically
indistinguishable.

\section{The Statistical Properties of the Correlations} \label{sec:stats}

To trace the origin of the correlations, we create two new bands with
only 32 modes each for the 200~\hMpc box simulation. One band is centred at
$k=0.47$~\Mpch and the other at $k=0.71$~\Mpch.
We have first verified that the density fluctuations $\delta_k$ themselves
are uncorrelated, as they must be by homogeneity.  We find  $W=0.008$, with
a probability of non-correlation of $p=0.44$.
For the two band-averaged $P(k)$ the correlation is $\rho=0.23$,
with a probability $p=6.3\times10^{-9}$ that the two are uncorrelated.
For the full set of $(64\times63)/2$ $P(k)$ correlations, we find
$W=0.025$, with a probability $p=2.3\times10^{-22}$, demonstrating the
clear presence of highly significant correlations among the individual $P(k)$.
However, we find that none of the values in the $32\times32$
inter-band correlation matrix for the individual $P(k)$ estimates is as great
as the correlation value $\rho=0.23$ found between the band-averaged power
spectra above. Indeed, the average correlation value in the matrix is $0.0099$,
and the standard deviation $0.037$. We conclude that
{\it the strong correlations among the band-averaged power spectra are built
up in the band-averaging procedure itself from the much weaker individual mode
correlations.}

We may understand the origin of the correlations by considering
a thin shell of width $\Delta k$ centred at $k_i$. Neglecting the
shot-noise terms, the variance in the band-averaged power $P_i(k)$ is
${\rm var}[P_i(k)]=2P_i(k)^2/N_k+\overline{T}/V$, where $\overline{T}$
denotes an average over the configurations in the $k-$shell, and
$N_k=V(2\pi)^{-3}4\pi k^2\Delta k$ is the number of modes in the shell.
When the contribution of the tri-spectrum to the variance is negligible, we
find that the correlation in the band-averaged power is given by
$\rho[P_i(k),P_j(k')]\sim[\overline{T}/ P_i(k)P_j(k')] kk'\Delta k$,
and $V$ has explicitly cancelled.
Here $\overline{T}$ denotes the tri-spectrum averaged over all
the configurations between both $k-$shells. For logarithmically spaced
frequencies this becomes
$\rho\sim[\overline{T}/P_i(k)P_j(k')] (kk')^{3/2}\Delta\log k$.

To make further progress, let us assume the hierarchical clustering
{\it ansatz}.  The form of the reduced 4-point function $\eta$ is then
$\eta(\bx_1,\bx_2,\bx_3,\bx_4)=R_a[\xi(\bx_{12})\xi(\bx_{23})\xi(\bx_{34}) +
{\rm cyc.}\ (12 {\rm terms})]+R_b[\xi(\bx_{12})\xi(\bx_{13})\xi(\bx_{14}) +
{\rm cyc.}\ (4 {\rm terms})]$, where $\bx_{ij}=\bx_i-\bx_j$ and $R_a$ and
$R_b$ are constants of order one \cite{P80,Fry84}.
Using equation~(\ref{eqn:trispectrum}), we obtain for the contribution to
the off-diagonal power spectrum covariance
\begin{eqnarray}
 && V^{-1}T(\bk,-\bk,\bkp,-\bkp)= \label{eqn:Tha} \\
 && \frac{R_a}{V}\left[P(|\bk-\bkp|)+ P(|\bk+\bkp|)\right]
    \left[P(k)+P(k')\right]^2 + \nonumber \\
 && \frac{R_b}{V}P(k)P(k')\left[P(k)+P(k')\right]. \nonumber
\end{eqnarray}
We find then that the power correlations for individual modes of frequencies
$k$ and $k'$ will be on the order of $[P(k)+P(k')]/V\sim10^{-3}-10^{-2}$ for
a 200\hMpc box. After band-averaging, and for $P(k')\gg P(k)$, the correlations
become $\rho\sim P_j(k')kk'\Delta k$ for linearly spaced $k-$shells, and
$\rho\sim\Delta_j^2(k')(k/k')^{3/2}\Delta\log k$ for logarithmic spacing.
Thus for logarithmically spaced shells, at a fixed $k'$ the correlations will
increase with $k$ like $k^{3/2}$. When $\Delta_i^2(k)\sim (\Delta\log k)^{-1}$,
the tri-spectrum term will
dominate the variance of $P_i$ and the correlations will flatten to
$\rho\sim[\Delta_i^2(k)]^{-1/2}\Delta_j^2(k')(k/k')^{3/2}(\Delta\log k)^{1/2}$.
This is just the behaviour displayed in Figure~\ref{fig:crz}. Motivated by
the hierarchical {\it ansatz} form, we find the correlations for $k\gg k'$ and
$P(k')\gg P(k)$ are well approximated by
\begin{equation}
\rho\approx \left(\overline{R}_a + \overline{R}_b\right)
\left(\frac{k}{k'}\right)^{3/2}\frac{\Delta^2(k')\Delta\log k}
{1+\overline{R}_b\left[\Delta^2(k)\Delta\log k\right]^{1/2}},
\end{equation}
where $\overline{R}_a=(34/21)^2$ and $\overline{R}_b=682/189$ are
angle-averaged coefficients derived in perturbation theory \cite{Fry84}. We
find that a direct computation using equation~(\ref{eqn:Tha}) with
$\overline{R}_a$ and $\overline{R}_b$ agrees less well with the simulations,
over-estimating the correlations by as much as a factor of 2. It is unclear
why the direct computation fails, but it improves toward the higher $k$ shells,
which may indicate that using the angle-averaged coefficients is an inadequate
approximation for the bins with fewer modes. Since performing the full set of
integrals in perturbation theory for all the modes in the $k-$shells is
extremely cumbersome, we will not pursue the hierarchical {\it ansatz} any
further, but note that it may be worth greater consideration.

A consequence of the correlations is a reduction in the number of degrees of
freedom describing the variance of the power spectrum.
While the distribution of {\it band-averaged\/} power is still $\chi^2$,
the modes are no longer independent and the ``effective number'' of degrees
of freedom will be less than the number of modes in the band.
We may define the ``effective number'' of degrees of freedom $n_{\rm dof}$ by
\begin{equation}
  n_{\rm dof}=\frac{2\langle\widehat{P}\rangle^2}{{\rm var}(\widehat{P})}.
\end{equation}

\begin{figure}
\begin{center}
\leavevmode \epsfxsize=3.3in \epsfbox{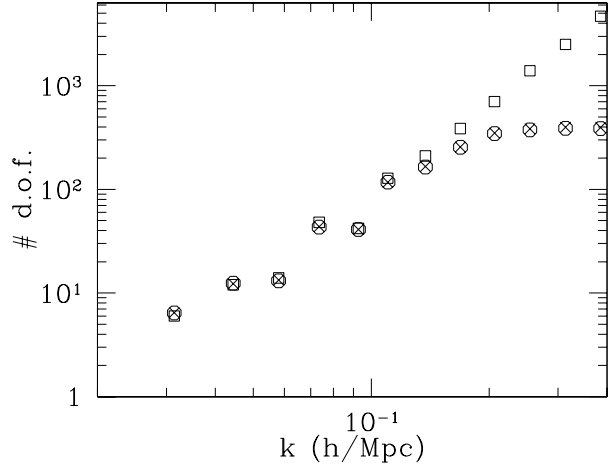}
\end{center}
\caption{The number of degrees-of-freedom expected in each frequency band based
on counting modes in the simulation box (squares) vs the equivalent number
$n_{\rm dof}$ found in the simulations (circles). Also shown are
maximum-likelihood estimates of $n_{\rm dof}$ given by fitting the power
spectra to a $\chi^2$ distribution (crosses). Correlations between modes
reduce the effective number of degrees-of-freedom at high frequencies, and so
increase the dispersion in the band-averaged power spectrum estimates over the
expectation for uncorrelated modes.}
\label{fig:cdof}
\end{figure}

Since the density fluctuation $\delta_k$ for each mode consists of independent
real and imaginary parts, when the modes are Gaussian-distributed each mode will
contribute two degrees of freedom to the
average. The band-averaged power spectrum will then be distributed like
$\chi^2$ with $n_{\rm dof}$ degrees of freedom. Correlations between modes will
reduce $n_{\rm dof}$ below the number expected from mode counting. This is a
strong effect at high frequencies, as shown in Figure~\ref{fig:cdof}. Motivated
by the hierarchical {\it ansatz}, we find that the reduction factor is
approximately described by $1+2\overline{R}_b\Delta^2(k)\Delta\log k$.
While $n_{\rm dof}$ agrees with the number of degrees of freedom expected for
Gaussian-distributed modes in the linear to quasi-linear regimes
($k<0.1$~\Mpch), on scales for which the fluctuations are non-linear
$n_{\rm dof}$ falls substantially short.
The errors on the power spectrum in the non-linear regime will correspondingly
greatly exceed estimates based on the assumption of Gaussian statistics.

In spite of the presence of strong correlations, the distribution of
band-averaged power spectra is still reasonably well-described by a $\chi^2$
distribution.
We show the maximum-likelihood estimates for the numbers of degrees of freedom
in Fig.~\ref{fig:cdof} found by assuming the power spectra in a given band are
distributed like $\chi^2$.
Formally, the KS test rejects the $\chi^2$ distribution with 90--99.7\%
confidence at frequencies in the range $0.25<k<2$~\Mpch. (We
haven't examined higher frequencies.) The distribution of the redshift-space
estimated power spectra, however, is found to be very close to $\chi^2$
at all frequencies.

\section{Conclusions} \label{sec:conclusions}

Estimates of the matter power spectrum from galaxy surveys have
generally assumed the power in separate modes to be uncorrelated. The
gravitational growth of the modes, however, will give rise to a tri-spectrum
which will induce correlations in the band-averaged power spectrum. In a set
of numerical experiments, we find that significant correlations in
power develop when at least one of the modes is non-linear, and approaching
100\% when the modes in both bands are non-linear.  Although the power spectrum
correlations are weaker in redshift space, they are still
statistically significant.

In addition to redshift space distortions, the measured galaxy power spectrum
will also depend on the bias between the galaxy count fluctuations and the
underlying matter fluctuations.
We have not computed the correlations in power allowing for bias.
Although this is straightforward within any given bias scheme, which scheme,
if any, the true galaxy distribution follows is unclear at this time.
More direct measurements of the dark matter fluctuations may be obtained
through galaxy velocity flows \cite{Dekel,StrWil} or lensing
\cite{Blandford,ME,K92,K98,Sel98,BerWaeMel,JaiSelWhi}.
Our results are not directly applicable to these measures, as they involve
projections (integrals) of the matter power spectrum. However, we expect that
measurement errors will again exceed those estimated assuming
Gaussian-distributed density fluctuations as a result of correlations in the
matter power spectrum.

The presence of power spectrum correlations restricts the range of
applicability of many techniques for fitting models to the measured power
spectra (e.g. Tegmark et al. 1998) to very large scales. In particular, it may
no longer be assumed that the {\bf $k$}-space density fluctuations
$\delta_k$ are distributed as a Gaussian random process:\ the tri-spectrum
must explicitly be taken into account if {\it any\/} of the modes fit are in
the non-linear regime. This constraint may be particularly severe if the
window function of the survey is broad, mixing modes across a wide range of
scales, and/or the survey is not very deep.
Correlations will affect the model-fitting based on all completed galaxy
surveys, and may only be circumvented with wide-angle deep surveys like the
Sloan Digital Sky Survey and the 2dF survey now underway.
For a catalog the depth of the SDSS Northern Polar Cap redshift survey,
the minimum frequency spacing for uncorrelated bins is
$\Delta k\approx0.015$~\Mpch \cite{MWP99}. On a scale of
$k=0.19$~\Mpch, the scale of non-linearity for the model investigated here,
the expected correlation with the power at $k'=0.1$~\Mpch will be on the order
of $\rho\approx0.15$.
Fitting a given model to measurements of the power spectrum in the non-linear
regime will require computing the correlations for the model. Provided there is
a sufficient number of modes per band, the Central Limit Theorem guarantees
the band-averaged power spectrum estimates will be distributed as a multivariate
Gaussian random process, simplifying the fitting process. Designing an efficient
fitting procedure in the presence of correlations is a topic worthy of future
investigation.

\bigskip
The authors would like to thank S. Colombi, J. Fry, and A. Heavens for helpful
discussions. We note that Scoccimarro, Zaldarriaga, \& Hui (1999)
independently recognized the existence of power spectrum correlations.
M.W. was supported by the NSF.


\begin{thebibliography}{99}
\bibitem[\protect\citename{Bernardeau, van Waebeke \& Mellier }1997]{BerWaeMel}
Bernardeau F., van Waebeke L., Mellier Y., 1997, Astron. Astrophys., 322, 1
\bibitem[\protect\citename{Bertschinger }1995]{B95}
Bertschinger E., 1995, preprint, astro-ph/9506070
\bibitem[\protect\citename{Blandford et al. }1991]{Blandford}
Bland	ford R.D., Saust A.B., Brainerd T.G., Villumsen J.V., 1991,
  MNRAS, 251, 600
\bibitem[\protect\citename{Bouchet et al. }1993]{BSDFYH}
Bouchet F., Strauss M., Davis M., Fisher K., Yahil A., Huchra J., 1993,
  ApJ, 417, 36
\bibitem[\protect\citename{Bunn \& White }1997]{BW97}
Bunn E.~F. \& White M., 1997, ApJ, 480, 6
\bibitem[\protect\citename{Colley }1997]{Colley}
Colley W.N., 1997, ApJ, 489, 471
\bibitem[\protect\citename{Dekel }1994]{Dekel}
Dekel A., 1994, ARA\&A, 32, 371
\bibitem[\protect\citename{Fan \& Bardeen }1995]{FB95}
Fan Z., Bardeen J.~M., 1995, Phys. Rev. D, 51, 6714
\bibitem[\protect\citename{Feldman, Kaiser \& Peacock }1994]{FKP}
Feldman H., Kaiser N., Peacock J., 1994, ApJ, 426, 23
\bibitem[\protect\citename{Fry }1984]{Fry84}
Fry J., 1984, ApJ, 279, 499
\bibitem[\protect\citename{Gazta\~naga }1994]{Gaz}
Gazta\~naga E., 1994, MNRAS, 268, 913
\bibitem[\protect\citename{Heavens }1998]{Heavens}
Heavens A.F., 1998, preprint, astro-ph/9804222
\bibitem[\protect\citename{Jain, Seljak \& White }1998]{JaiSelWhi}
Jain B., Seljak U., White S.D.M., 1998, preprint [astro-ph/9804238]
\bibitem[\protect\citename{Juszkiewicz }1981]{Jus81}
Juszkiewicz R., 1981, MNRAS, 197, 931
\bibitem[\protect\citename{Kaiser }1992]{K92}
Kaiser N., 1992, ApJ, 388, 272
\bibitem[\protect\citename{Kaiser }1998]{K98}
Kaiser N., 1998, ApJ, 498, 26
\bibitem[\protect\citename{Kogut et al. }1996]{KBBGHSW}
Kogut A., et al., 1996, ApJ, 464, L29
\bibitem[\protect\citename{Meiksin et al. }1999]{MWP99}
Meiksin A., White M., Peacock J., 1999, MNRAS, in press [astro-ph/9812214]
\bibitem[\protect\citename{Miralda-Escude }1991]{ME}
Miralda-Escude J., 1991, ApJ, 380, 1
\bibitem[\protect\citename{Nusser, Dekel \& Yahil }1994]{NusDekYah}
Nusser A., Dekel A., Yahil A., 1994, ApJ, 449, 439
\bibitem[\protect\citename{Peebles }1980]{P80}
Peebles P.J.E., 1980, The Large-Scale Structure of the Universe.
Princeton Univ. Press, Princeton
\bibitem[\protect\citename{Press et al. }1992]{NumRec}
Press W.H., Teukolsky, S.A, Vetterling, W.T., Flannery, B.P., 1992,
Numerical Recipes, 2nd edn. Cambridge Univ. Press, Cambridge
\bibitem[\protect\citename{Scoccimarro, Zaldarriaga \& Hui }1999]{Sco99}
Scoccimarro R., Zaldarriaga M., Hui L., 1999, submitted to ApJ
[astro-ph/9901099]
\bibitem[\protect\citename{Seljak }1998]{Sel98}
Seljak U., 1998, ApJ, 506, 64
\bibitem[\protect\citename{Siegel \& Castellan }1985]{NPS}
Siegel S., Castellan N.S. Jr, 1985, Nonparametric Statistics.
McGraw Hill, New York
\bibitem[\protect\citename{Strauss \& Willick }1995]{StrWil}
Strauss M.A., Willick J.A., 1995, Phys. Rep. 261, 271
\bibitem[\protect\citename{Tegmark et al. }1998]{Teg98}
Tegmark M., Hamilton A., Strauss M., Vogeley M., Szalay A., 1998,
ApJ, 499, 555
\bibitem[\protect\citename{Vishniac }1983]{Vish83}
Vishniac E.~T., 1983, MNRAS, 203, 345
\bibitem[\protect\citename{White }1998]{Moments}
White M., submitted to MNRAS, [astro-ph/9811227]
\bibitem[\protect\citename{White, Efstathiou \& Frenk }1993]{WEF93}
White S.D.M., Efstathiou G.P., Frenk C.S., 1993, MNRAS, 262, 1023
\end{thebibliography}
\end{document}